# Coupled pyroelectric-photovoltaic effect in 2D ferroelectric $\alpha$-In$_2$Se$_3$.

Michael Uzhansky, Abhishek Rakshit, Yoav Kalcheim and Elad Koren[*]

*Faculty of Materials Science and Engineering, Technion - Israel Institute of Technology, Haifa, 3200003, Israel.*
[*]Email – eladk@technion.ac.il

## Abstract-

Pyroelectric and photovoltaic effects are vital in cutting-edge thermal imaging, infrared sensors, thermal and solar energy harvesting. Recent advances revealed the great potential of the bulk photovoltaic effect in two-dimensional (2D) semiconductor-ferroelectric materials to enable reconfigurable *p-n* junction operation with the potential to surpass the Shockley-Queiseer limit. Moreover, the extremely low thickness, high thermal conductivity, dangling bonds free interface, and room-temperature stable ferroelectricity down to a single monolayer endow 2D ferroelectrics with a superior pyroelectric figure of merit. Herein, we performed direct pyroelectric measurements of 2D $\alpha$-In$_2$Se$_3$ under dark and light conditions. The results reveal a gigantic pyroelectric coefficient of ~ 30.7 mC/m$^2$K and a figure of merit of ~135.9 m$^2$/C. In addition, we perform temperature-dependent short-circuit photovoltaic response measurements in which the excess photocurrent is modulated in proportion with the temperature variations due to the induced in-plane potential variations. Consequently, the discovered pyroelectric-photovoltaic effect allows the combination of direct temperature (photovoltaic) and temperature-derivative (pyroelectric) sensing. Finally, we utilized the intercoupled ferroelectricity of In$_2$Se$_3$ to realize a non-volatile, self-powered photovoltaic memory operation, demonstrating a stable short-circuit current switching with a decent 10$^3$ ON-OFF ratio. The coupled pyroelectric-photovoltaic effect, along with reconfigurable photocurrent, pave the way for a novel monolithic device technology with integrated thermal and optical response, in-memory logic and energy harvesting.

## Introduction-

Ferroelectric materials have been changing the paradigm of contemporary science and engineering since Valasek's first discovery in 1920[1]. Such immense importance stems from their outstanding multifunctionality for non-volatile memory[2], field-effect transistors (FET)[3,4], actuators[5], transducers[6], photovoltaics[7,8], etc. The pyroelectric effect describes the change in spontaneous polarization with temperature variations, where the net change in dipole moment introduces capacitive current flow that is proportional to the temperature derivative with time dT/dt. This effect is of utmost importance for self-powered infrared thermal-sensing[9] and waste heat-energy harvesting[10]. While the nanoelectronics industry demands the utter miniaturization of the ferroelectrics films, the ultra-low thickness and, hence, low heat capacity entails an extremely small thermal time constant, which is highly beneficial for pyroelectric responsivity[11]. However, conventional ferroelectrics (CFs) such as perovskite oxides ($ABO_3$) do not often meet such requirements at the nanoscale due to the sharp reduction in spontaneous polarization caused by a strong depolarization field, interfacial defects due to surface reconstruction, electron screening, etc[12,13].

Two-dimensional (2D) materials have been drawing utterly large attention during the past few decades after the first experimental realization of graphene[14]. Their outstanding and diverse properties such as large surface-to-volume ratio, tunable bandgap[15–17], ultra-strong light-matter interaction[18], excellent thermal and electrical conductivity[19,20] etc., drive the constantly growing interest. The recently reported room-temperature stable ferroelectricity in monolayer-thin 2D materials such as $\alpha$, $\beta$-$In_2Se_3$[21–23], $CuInP_2S_6$ (CIPS)[24], SnTe[25], and $MoTe_2$[26] possess dangling-bonds and reconstruction free interfaces, can address the scaling shortcomings of CFs. In addition, the existing bulk photovoltaic effect in 2D ferroelectrics possesses the potential to overcome the Shockley-Queiseer limit in conventional photovoltaics [27,28].

Among 2D ferroelectrics, $\alpha$-$In_2Se_3$ possesses a unique inter-coupled in-plane (IP) and out-of-plane (OOP) polarizations [21,29–33], allowing to control the horizontal ferroelectric field via an applied vertical bias and vice versa. In contrast with CFs, which are insulators, $In_2Se_3$ is an *n*-type semiconductor with a direct bandgap of ~1.4 eV[17] that allows its embedding as the channel material of FET and optical sensors[29,34]. Moreover, recent molecular dynamic simulations revealed a high Curie temperature of ~ 650 K in $\alpha$-$In_2Se_3$ [35], enabling a wide operational range. These unique features have found realization in novel optoelectronic memories [29,36,37], reconfigurable *p-n* junctions[33,38], and neuromorphic computing.[39–41]

Herein, we demonstrate direct pyroelectric measurement of back-gated FET based on *α*-In$_2$Se$_3$ under light and dark conditions. The extracted giant pyroelectric coefficient of ~30.7 mC/m$^2$K and figure of merit of ~135.9 m$^2$/C reveal a supreme thermal-sensing performance, significantly surpassing CFs[42]. Under light illumination, the built-in IP polarization and the semiconducting nature of *α*-In$_2$Se$_3$ introduce the bulk photovoltaic effect, which enables the realization of self-powered and energy-harvesting devices without the need for complex heterostructures fabrication for constructing *p-n* junctions. Intriguingly, we reveal a coupled pyroelectric-photovoltaic effect where the short-circuit current magnitude linearly follows the temperature modulation while the pyroelectric current follows the temperature derivative. Finally, the back-gate electrode was used to control the IP polarization based on the intercoupled ferroelectricity in In$_2$Se$_3$, allowing to switch the short-circuit current polarity in a non-volatile fashion. The results reveal the extreme temperature-sensitive pyroelectric and photovoltaic effects enabling the detection of temperature and temperature-derivative modulations, respectively by the same device. The technology paves the way for integrated monolithic photovoltaic-pyroelectric elements for coupled thermal-sensing and optoelectronic in-memory applications.

## Results –

  $\alpha$-In$_2$Se$_3$ flakes were mechanically exfoliated onto the $p^{++}$-Si substrate covered with 300 nm of SiO$_2$. Source and drain contacts were fabricated using standard electron beam lithography with subsequent metal deposition of 5/50 nm thick Cr/Au metals. Figure (S1) in Supplementary Information depicts the AFM topography and Raman spectroscopy analysis. Figure 1a illustrates the experimental setup where the back-gated field effect transistor (FET) was placed onto a programmable heater stage. The source and drain contacts were grounded throughout the measurements, and a broad-band white LED source of spectral range 420–720 nm and intensity of ~332 µW/cm$^2$ was used for device illumination. Figure 1b presents a schematic illustration of the energy band structure and current generation of the pyroelectric-photovoltaic In$_2$Se$_3$ device under light illumination and temperature variations. In particular, temperature variations under dark conditions modify the net dipole moment, which is followed by capacitive current flow to compensate for the change in bound charges. During illumination, the inherent IP ferroelectric field separates the photogenerated electron-hole pairs within the In$_2$Se$_3$ semiconductor channel and creates a short-circuit current. The ferroelectric polarization reduces once the device temperature increases, and the photocurrent monotonously drops. Thus, while the pyroelectric current follows the temperature derivative, the short-circuit current follows the temperature modulation itself. Figure 1c presents the measured polar plot of the normal-incident angular-dependent second-harmonic generation (SHG), which indicates the 6-fold inversion-symmetry breaking of the IP polarization in In$_2$Se$_3$. Additional polar plots measured at different locations across the device are included in the Supplementary information (Fig. S2). Domain structure homogeneity was evaluated by performing SHG mapping (Fig. 1d). The optical image of the studied device is depicted in the inset of Figure 1d. The observed local variations in SHG intensity indicate that some domains are better aligned along the 90-degree polarization axis. Likewise, the darker regions may consist of domains with partly anti-parallel polarization directions, which eventually cancel each other, resulting in a smaller response.[43]

  Direct pyroelectric measurement under dark conditions was conducted to retrieve the pyroelectric coefficient and figure of merit (Fig. 2a). The top panel shows the temperature modulation throughout the experiment. In particular, the device was repeatedly heated to 40 °C followed by natural cooling down to 30 °C under ambient conditions. The corresponding temperature derivative dT/dt and the pyroelectric current ($I_{pyro}$) are shown in the middle and bottom panels, respectively. Evidently, the measured current clearly follows the temperature derivative, confirming the existence of the pyroelectric effect. The device shows stable performance throughout multiple temperature cycles without current degradation. It is important to note that the current does not change its sign during the cooling cycles. This can be attributed both to the substantially smaller temperature derivative during cooling and to the semiconducting nature of In$_2$Se$_3$, which facilitates free-charge transport. In particular, the thermal generation of electron-hole pairs and their consequent separation by the IP ferroelectric field results in a non-zero-base current, which screens the negative pyroelectric current during cooling. Similarly, the induced photocurrent due to the presence of free charge carriers under light illumination conditions shows linear dependence with temperature modulation, as will be demonstrated in the following section. The pyroelectric coefficient $p \sim 30.7$ mC/m$^2$K and the figure of merit $F_v \sim 135.9$ m$^2$/C were calculated using the following equations:

$$p = \frac{I_{pyro}}{A \cdot \frac{dT}{dt}} \quad (1)$$

$$F_v = \frac{p}{c_p \varepsilon_r \varepsilon_0} \quad (2)$$

where, $I_{pyro}$ is the pyroelectric current (the difference between the current maximum and the base values), $A$ – the surface area of the device, $T$ - temperature, $c_p$ – specific heat capacity (for In$_2$Se$_3$ is 1.5 J K$^{-1}$cm$^{-3}$)[44], $\varepsilon_0$ – permittivity of the free space and $\varepsilon_r$ – relative permittivity (~17 for In$_2$Se$_3$)[45]. Figure 2b presents the extracted parameters of this work in comparison with several different CFs (PZT, PVDF, LaTiO$_3$)[46–48] and recently reported 2D materials and nanomembranes (ZnO, β-In$_2$Se$_3$).[49] The extracted giant pyroelectric coefficient and the figure of merit of α-In$_2$Se$_3$ significantly surpass previously reported values for CFs and 2D materials.

Figure 3a presents the measured current (black curve) and temperature (red curve) versus time under dark and light (designated with yellow box) conditions. The device was cyclically heated to ~40 °C and subsequently naturally cooled to ~30 °C under ambient conditions. While the pyroelectric effect is prominent in dark conditions (manifested by the current peak of ~160 pA during the highest temperature change), illumination of the device generates a substantial short-circuit current (~1 nA), which effectively screens the pyroelectric effect. In addition, the photocurrent linearly follows the temperature instead of its derivative with time. The short-circuit current reduces with temperature rise due to the corresponding decreases in the IP polarization field. Similarly, the subsequent cooling increases the IP electric field, which restores the previous current level. Importantly, the short-circuit current demonstrates a great temperature sensitivity of ~ 20 pA/°C within the 30-40 °C range. The distinct response under dark and light conditions is further illustrated in Figures 3b (i, ii) by the enlarged sections from Figure 3a. In particular, while the pyroelectric current dominates under dark conditions by following the temperature derivative (Fig. 3bi), the photocurrent monotonously follows the temperature (Fig. 3bii). This feature presents a novel combination of thermal and photo-sensing by a coupled pyroelectric-photovoltaic effect within the same material. Consequently, the device allows to directly probe temperature or its derivative without the need to apply a bias potential across the device. Next, the intercoupled in-plane (IP) and out-of-plane (OOP) ferroelectricity in $\alpha$-In$_2$Se$_3$ was used to realize self-powered optoelectronic memory operation based on bulk photovoltaic effect. This unique coupling allows the control of the horizontal polarization by an external vertical field stimulus and vice versa. Hence, 30-second voltage pulses of -100V (blue curve) and +100V (red curve) were applied to the back-gate electrode to switch the IP polarization direction (Fig. 3c). During the measurements, the source, drain, and gate electrodes were grounded, and a broadband white light source was used to generate multiple 10-second light pulses to generate the short-circuit current. The switching of the IP field results in two opposite current levels, which can be used as "1" and "-1" in memory or logic applications or for weights with opposite signs, which are attractive for neuromorphic-based image processing[50]. The device shows a stable short-circuit current generation under several cycles of light pulses without current-level degradation.

Figures 4a and 4b present the temperature dependent pyroelectric and short-circuit current device characteristics, respectively. The device was gradually heated by 5 °C in two-minute steps from 25 to 60 °C. Figure 4a shows the measurement in dark conditions, where a strong pyroelectric current is generated precisely during the temperature modulation and monotonously decreases with temperature increase. The inset image in Figure 4a describes a linear correlation between the peak current magnitude and temperature. The base current level at dark caused by thermal generation and subsequent charge separation of electron-hole pairs by the IP polarization reduces with the increase in temperature due to the decrease in the IP field. Under light illumination, the photocurrent decreases from ~1.5 nA at room temperature to ~0.5 nA at 65 °C, indicating a substantial temperature sensitivity (Fig. 4b). In both cases, the IP polarization field reduces with increasing temperature, resulting in weaker pyroelectric and photovoltaic effects. Figure S3 in the SI section presents a similar measurement that includes both heating and cooling cycles under light and dark conditions. The results show identical current levels prior to the heating and after the cooling cycles, confirming the operation stability. Importantly, the fully restored dark and photocurrent levels upon cooling indicate that the major contribution to the observed current modulations is due to induced changes in the IP field and not due to trapping/de-trapping of fixed charges at the $In_2Se_3$ itself or at its interface with the $SiO_2$ support.

The direct pyroelectric measurements performed at ambient conditions (Figure 2a) possess a slow natural cooling rate, resulting in a minute pyroelectric current. To overcome this, direct pyroelectric characterization was performed in a cryogenic environment using liquid $N_2$ to enhance the cooling power. Starting at a sample temperature of 270 K we were able to generate cooling rates of 10 – 40 K/min (Figure S4). These cooling rates are ~5-20 times higher than at ambient conditions allowing for measurable pyroelectric current in the opposite direction to the current generated during heating. Figure 4c demonstrates the linear relation between the pyroelectric current and temperature modulation rate for both cooling and heating, which agrees with Equation (1).

Another device of a similar geometry (see Supplementary Information Fig. S5 for the optical and AFM images) was used to study the temperature dependence of the transfer ($I_D$-$V_G$) and output ($I_D$-$V_D$) electrical characteristics (Fig. 4d). The temperature range (from room temperature (RT) to 200 °C) was set below the $\alpha \to \beta$ $In_2Se_3$ phase transition, which occurs above 250 °C[49,51]. Raman spectra showing $\alpha \to \beta$ $In_2Se_3$ phase transition above 250 °C can be found in Supplementary Information (Fig. S6). Figure 4c presents the transfer characteristics in dark for a drain bias of 1 V. The wide clockwise hysteresis loop at RT manifests the ferroelectricity in $\alpha$-$In_2Se_3$. As the temperature increases, the hysteresis loop gradually decreases in conjugation with a distinct reduction in the on-off current levels ratio. These variations are attributed to 1) the reduction in ferroelectric polarization with increasing temperature and 2) the increase in charge carrier concentration due to thermal generation, which also acts to screen the ferroelectric polarization field. The device output and transfer characteristics at room temperature in dark and light conditions can be found in the SI (Fig. S7).

## Conclusions –

Direct pyroelectric measurements of 2D $\alpha$-In$_2$Se$_3$-based FET were performed to extract pyroelectric coefficient of ~30.7 mC/m$^2$K and a figure of merit of ~135.9 m$^2$/C, indicating a superior thermal-sensing performance. In addition, a coupled pyroelectric-photovoltaic effect was demonstrated, where the pyroelectric current follows the temperature derivative, while the short-circuit current monotonously follows temperature. Moreover, intercoupled ferroelectricity was utilized to realize a novel self-powered optoelectronic memory device based on the bulk photovoltaic effect and the gate-induced modulation in IP polarization. The fabricated device demonstrated a decent $10^3$ ON-OFF ratio at room temperature and stable short-circuit current direction switching. To this end, the presented device features pave the way for advanced monolithically integrated self-powered thermal-photo sensors and memories and energy harvesting devices.

## Methods –

### Sample preparation -

α-In$_2$Se$_3$ crystals (2D Semiconductors, 99.9999% purity) were mechanically exfoliated onto degenerately doped $p^{++}$-type silicon substrate encapsulated with thermally grown 300 nm SiO$_2$. The first few layers of the bulk crystals were discarded to prevent contamination and the subsequent pristine α-In$_2$Se$_3$ flakes were used for device fabrication.

### Device fabrication –

The source and drain metal electrodes (Cr/Au 5/60 nm) were fabricated using electron beam lithography (Raith-eLine) followed by electron beam evaporation (Evatec BAK-501A). Before metal deposition, a short oxygen plasma treatment was used to eliminate undesired resist residuals. The metal deposition rate was 0.5 Å/s for Cr and 1 Å/s for Au at a chamber pressure of around $7 \cdot 10^{-7}$ torr.

### Surface and optoelectronic characterization –

Atomic force microscopy (AFM) in tapping mode (Dimension-Scanassist Bruker Inc.) was used to study the device topography in an N$_2$-filled glovebox (H$_2$O and O$_2$ content <1 ppm). Electrical measurements were performed using semiconductor device parameter analyzers Keysight B1500A, precision source/measure Keysight B2900A unit, and a probe station equipped with an optical microscope. A broadband white LED source of spectral range 420–720 nm and intensity of ~332 µW/cm$^2$ was used for short-circuit current measurements. A Linkman (HFS350EV-PB4) heating/cooling stage was used to control the device temperature at ambient conditions. *In-situ* device temperature was recorded using a K-type thermocouple placed on the Si-SiO$_2$ substrate carrying the device and was simultaneously sampled with short-circuit current by Keysight B1500A.

### Cryogenic electronic characterization –

Direct pyroelectric measurements at low temperature were conducted in Lakeshore TTPX cryogenic probe station under a high vacuum of ~ $4 \cdot 10^{-6}$ torr. The device temperature was controlled by the heating stage and liquid nitrogen at a temperature range of ~270K +/- 10K.

**Spectroscopic characterization –**

Raman spectroscopy (WITec Alpha300R) was performed in confocal mode using a 532 nm laser with 100x objective (NA = 0.9; Δλ~ 360 nm, 1800 g*mm$^{-1}$ grating). The power of the laser was kept around 1 mW to avoid undesirable sample heating and degradation. 1064 nm laser (Rainbow1064OEM) with 100x objective (NA = 0.9; Δλ~ 360 nm, 600 g*mm$^{-1}$ grating) in normal incidence mode was used for angular-dependent second-harmonic generation (SHG) measurement. During the SHG mapping, the polarizer was rotated 90 degrees about the analyzer.

## Associated Content –

**Supplementary Information**

## Author Information –

**Corresponding author –**

Elad Koren - *Faculty of Materials Science and Engineering, Technion - Israel Institute of Technology, Haifa, 3200003, Israel.*

orcid.org/0000-0001-7437-7124; Email – eladk@technion.ac.il

**Authors -**

Michael Uzhansky - *Faculty of Materials Science and Engineering, Technion - Israel Institute of Technology, Haifa, 3200003, Israel.*

*orcid.org/0000-0002-4959-9750*

Yoav Kalcheim - *Faculty of Materials Science and Engineering, Technion - Israel Institute of Technology, Haifa, 3200003, Israel.*

*orcid.org/0000-0002-1489-0505*

Abhishek Rakshit - *Faculty of Materials Science and Engineering, Technion - Israel Institute of Technology, Haifa, 3200003, Israel.*

*orcid.org/0000-0002-3208-1496*

**Author Contributions –**

M.U. performed the experimental work, and E.K. supervised the work. Y.K. and A.R. provided experimental support with cryogenic electronic characterization. All authors participated in the manuscript writing.

**Conflict of Interest -**

The authors declare no conflict of interest.

## Acknowledgments -

M.U. gratefully acknowledges the support of The KLA Excellence Fellowship. E.K. gratefully acknowledges the Israel Innovation Authority (Kamin), the Israel Department of Energy and the Ministry of Innovation, Science & Technology for financial assistance and the Micro & Nano Fabrication Unit (MNFU) for the nanofabrication facilities. We thank Prof. Eilam Yalon for fruitful discussions.

# References –

1. Valasek, J. Piezo-Electric and Allied Phenomena in Rochelle Salt. *Physical Review* **17**, 475–481 (1921).

2. Sun, H. *et al.* Nonvolatile ferroelectric domain wall memory integrated on silicon. *Nat Commun* **13**, 4332 (2022).

3. Baek, S. *et al.* Ferroelectric Field-Effect-Transistor Integrated with Ferroelectrics Heterostructure. *Advanced Science* **9**, (2022).

4. Khan, A. I., Keshavarzi, A. & Datta, S. The future of ferroelectric field-effect transistor technology. *Nat Electron* **3**, 588–597 (2020).

5. Liu, Y. *et al.* Electro-thermal actuation in percolative ferroelectric polymer nanocomposites. *Nat Mater* **22**, 873–879 (2023).

6. Cross, L. E. Ferroelectric materials for electromechanical transducer applications. *Mater Chem Phys* **43**, 108–115 (1996).

7. Huang, H. Ferroelectric photovoltaics. *Nat Photonics* **4**, 134–135 (2010).

8. Han, X., Ji, Y. & Yang, Y. Ferroelectric Photovoltaic Materials and Devices. *Adv Funct Mater* **32**, (2022).

9. Chowdhury, Z. I., Imtiaz, M. H., Azam, M. M., Sumi, M. R. A. & Nur, N. S. Design and implementation of Pyroelectric Infrared sensor based security system using microcontroller. in *IEEE Technology Students' Symposium* 1–5 (IEEE, 2011). doi:10.1109/TECHSYM.2011.5783853.

10. Thakre, A., Kumar, A., Song, H.-C., Jeong, D.-Y. & Ryu, J. Pyroelectric Energy Conversion and Its Applications—Flexible Energy Harvesters and Sensors. *Sensors* **19**, 2170 (2019).

11. Whatmore, R. W. & Ward, S. J. Pyroelectric infrared detectors and materials—A critical perspective. *J Appl Phys* **133**, (2023).

12. Fong, D. D. *et al.* Ferroelectricity in Ultrathin Perovskite Films. *Science (1979)* **304**, 1650–1653 (2004).

13. Junquera, J. & Ghosez, P. Critical thickness for ferroelectricity in perovskite ultrathin films. *Nature* **422**, 506–509 (2003).

14. Novoselov, K. S. *et al.* Electric Field Effect in Atomically Thin Carbon Films. *Science (1979)* **306**, 666–669 (2004).

15. Tang, H. *et al.* Tunable band gaps and optical absorption properties of bent MoS2 nanoribbons. *Sci Rep* **12**, 3008 (2022).

16. Cui, C. *et al.* Intercorrelated In-Plane and Out-of-Plane Ferroelectricity in Ultrathin Two-Dimensional Layered Semiconductor $In_2Se_3$. *Nano Lett* **18**, 1253–1258 (2018).

17. Lyu, F. *et al.* Thickness-dependent band gap of $\alpha$-$In_2Se_3$: from electron energy loss spectroscopy to density functional theory calculations. *Nanotechnology* **31**, 315711 (2020).

18. Zhang, T. & Qiu, C.-W. Light walking in low-symmetry 2D materials. *Photonics Insights* **2**, C03 (2023).

19. Chen, S. *et al.* Raman Measurements of Thermal Transport in Suspended Monolayer Graphene of Variable Sizes in Vacuum and Gaseous Environments. *ACS Nano* **5**, 321–328 (2011).

20. Lim, S., Park, H., Yamamoto, G., Lee, C. & Suk, J. W. Measurements of the Electrical Conductivity of Monolayer Graphene Flakes Using Conductive Atomic Force Microscopy. *Nanomaterials* **11**, 2575 (2021).

21. Ding, W. *et al.* Prediction of intrinsic two-dimensional ferroelectrics in In2Se3 and other III2-VI3 van der Waals materials. *Nat Commun* **8**, 14956 (2017).

22. Lee, S. *et al.* Low-temperature processed beta-phase $In_2Se_3$ ferroelectric semiconductor thin film transistors. *2d Mater* **9**, 025023 (2022).

23. Mukherjee, S. & Koren, E. Indium Selenide ($In_2Se_3$) – An Emerging Van-der-Waals Material for Photodetection and Non-Volatile Memory Applications. *Isr J Chem* **62**, (2022).

24. Zhao, M., Gou, G., Ding, X. & Sun, J. An ultrathin two-dimensional vertical ferroelectric tunneling junction based on $CuInP_2S_6$ monolayer. *Nanoscale* **12**, 12522–12530 (2020).

25. Chang, K. *et al.* Discovery of robust in-plane ferroelectricity in atomic-thick SnTe. *Science (1979)* **353**, 274–278 (2016).

26. Yuan, S. *et al.* Room-temperature ferroelectricity in MoTe2 down to the atomic monolayer limit. *Nat Commun* **10**, 1775 (2019).

27. Dai, Z. & Rappe, A. M. Recent progress in the theory of bulk photovoltaic effect. *Chemical Physics Reviews* **4**, (2023).

28. Li, Y. *et al.* Enhanced bulk photovoltaic effect in two-dimensional ferroelectric CuInP2S6. *Nat Commun* **12**, 5896 (2021).

29. Dutta, D., Mukherjee, S., Uzhansky, M. & Koren, E. Cross-field optoelectronic modulation via inter-coupled ferroelectricity in 2D In2Se3. *NPJ 2D Mater Appl* **5**, 81 (2021).

30. Li, Y., Gong, M. & Zeng, H. Atomically thin α-$In_2Se_3$: an emergent two-dimensional room temperature ferroelectric semiconductor. *Journal of Semiconductors* **40**, 061002 (2019).

31. Xue, F. *et al.* Gate-Tunable and Multidirection-Switchable Memristive Phenomena in a Van Der Waals Ferroelectric. *Advanced Materials* **31**, 1901300 (2019).

32. Wan, S., Peng, Q., Wu, Z. & Zhou, Y. Nonvolatile Ferroelectric Memory with Lateral β/α/β $In_2Se_3$ Heterojunctions. *ACS Appl Mater Interfaces* **14**, 25693–25700 (2022).

33. Dutta, D. *et al.* Edge-Based Two-Dimensional α-$In_2Se_3$–$MoS_2$ Ferroelectric Field Effect Device. *ACS Appl Mater Interfaces* **15**, 18505–18515 (2023).

34. Si, M. *et al.* A ferroelectric semiconductor field-effect transistor. *Nat Electron* **2**, 580–586 (2019).

35. Soleimani, M. & Pourfath, M. Ferroelectricity and phase transitions in $In_2Se_3$ van der Waals material. *Nanoscale* **12**, 22688–22697 (2020).

36. Mukherjee, S. *et al.* Scalable Integration of Coplanar Heterojunction Monolithic Devices on Two-Dimensional $In_2Se_3$. *ACS Nano* **14**, 17543–17553 (2020).

37. Mukherjee, S., Dutta, D., Uzhansky, M. & Koren, E. Monolithic In2Se3–In2O3 heterojunction for multibit non-volatile memory and logic operations using optoelectronic inputs. *NPJ 2D Mater Appl* **6**, 37 (2022).

38. Uzhansky, M., Mukherjee, S., Vijayan, G. & Koren, E. Non-Volatile Reconfigurable p–n Junction Utilizing In-Plane Ferroelectricity in 2D $WSe_2$/α-$In_2Se_3$ Asymmetric Heterostructures. *Adv Funct Mater* (2023) doi:10.1002/adfm.202306682.

39. Wang, L. *et al.* Exploring Ferroelectric Switching in α-$In_2Se_3$ for Neuromorphic Computing. *Adv Funct Mater* **30**, (2020).

40. Zhang, Y. *et al.* Analog and Digital Mode α-$In_2Se_3$ Memristive Devices for Neuromorphic and Memory Applications. *Adv Electron Mater* **7**, 2100609 (2021).

41. Mukherjee, S., Dutta, D., Ghosh, A. & Koren, E. Graphene-$In_2Se_3$ van der Waals Heterojunction Neuristor for Optical In-Memory Bimodal Operation. *ACS Nano* (2023) doi:10.1021/acsnano.3c03820.

42. Bauer, S. & Lang, S. B. Pyroelectric polymer electrets. *IEEE Transactions on Dielectrics and Electrical Insulation* **3**, 647–676 (1996).

43. Xiao, J. *et al.* Intrinsic Two-Dimensional Ferroelectricity with Dipole Locking. *Phys Rev Lett* **120**, 227601 (2018).

44. Boehnke, U.-C., Kühn, G., Berezovski≲, G. A. & Spassov, T. Some aspects of the thermal behaviour of In2Se3. *Journal of Thermal Analysis* **32**, 115–120 (1987).

45. Wu, D. *et al.* Thickness-Dependent Dielectric Constant of Few-Layer $In_2Se_3$ Nanoflakes. *Nano Lett* **15**, 8136–8140 (2015).

46. Putley, E. H. A method for evaluating the performance of pyroelectric detectors. *Infrared Phys* **20**, 139–147 (1980).

47. Whatmore, R. W., Osbond, P. C. & Shorrocks, N. M. Ferroelectric materials for thermal IR detectors. *Ferroelectrics* **76**, 351–367 (1987).

48. Ploss, B. & Bauer, S. Characterization of materials for integrated pyroelectric sensors. *Sens Actuators A Phys* **26**, 407–411 (1991).

49. Jiang, J. *et al.* Giant pyroelectricity in nanomembranes. *Nature* **607**, 480–485 (2022).

50. Cui, B. *et al.* Ferroelectric photosensor network: an advanced hardware solution to real-time machine vision. *Nat Commun* **13**, 1707 (2022).

51. Lyu, F. *et al.* Temperature-Driven α-β Phase Transformation and Enhanced Electronic Property of 2H α-$In_2Se_3$. *ACS Appl Mater Interfaces* **14**, 23637–23644 (2022).

# Figure 1

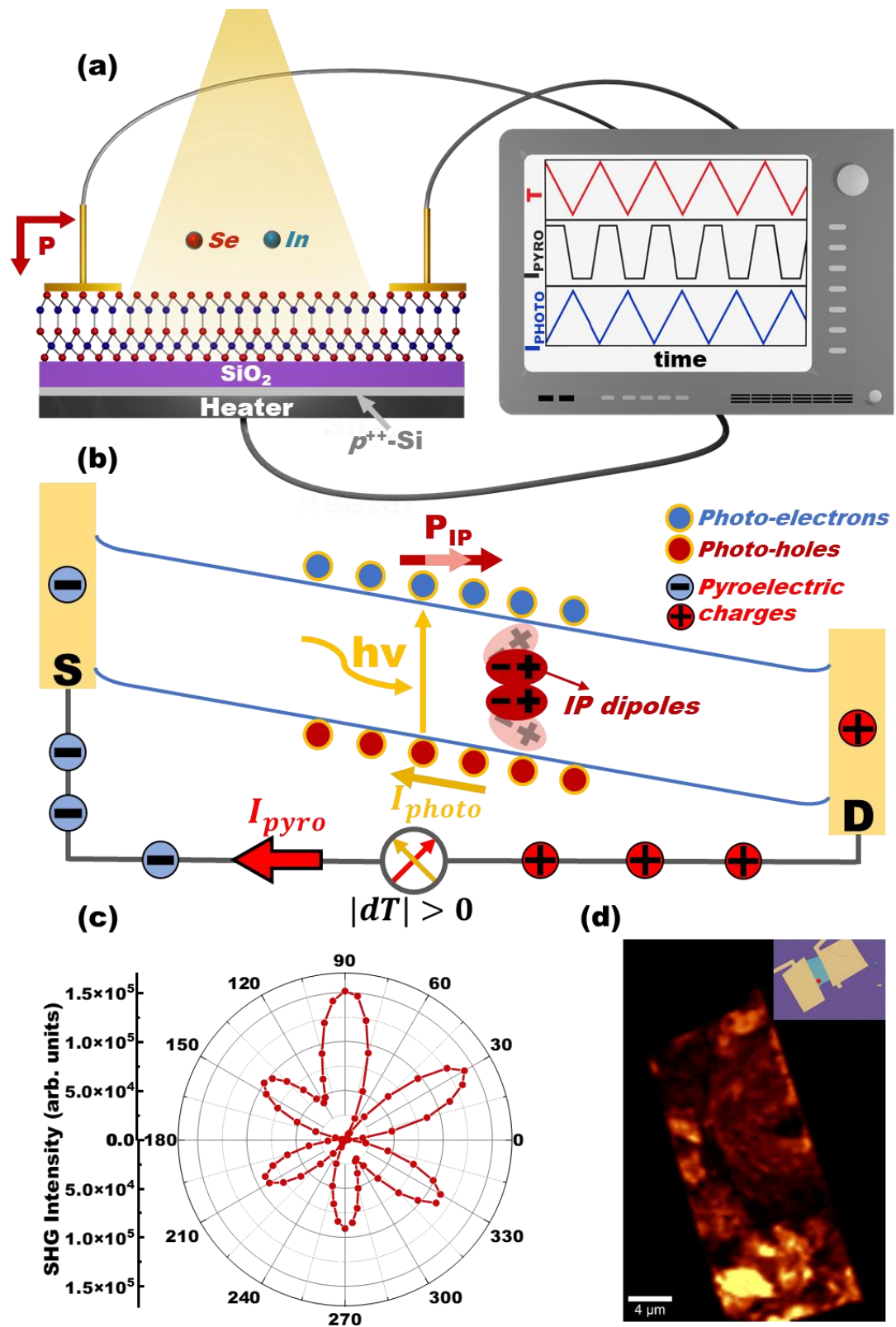

*Figure 1* – **(a)** Schematic representation of the α-In$_2$Se$_3$ based FET device and the pyroelectric-photovoltaic measurement setup. **(b)** – Schematic band structure representation of the coexisting pyroelectric and photovoltaic effects. **(c)** Polar plot of SHG intensity as a function of the excitation laser polarization **(d)** SHG mapping of the In$_2$Se$_3$ was measured for 90 degrees light polarization. Intensity variations reflect relative differences in the local dipole strength. The inset image presents the optical image of the device. Red point indicates the area where **(c)** was measured. Scale bar is 4 μm.

# Figure 2

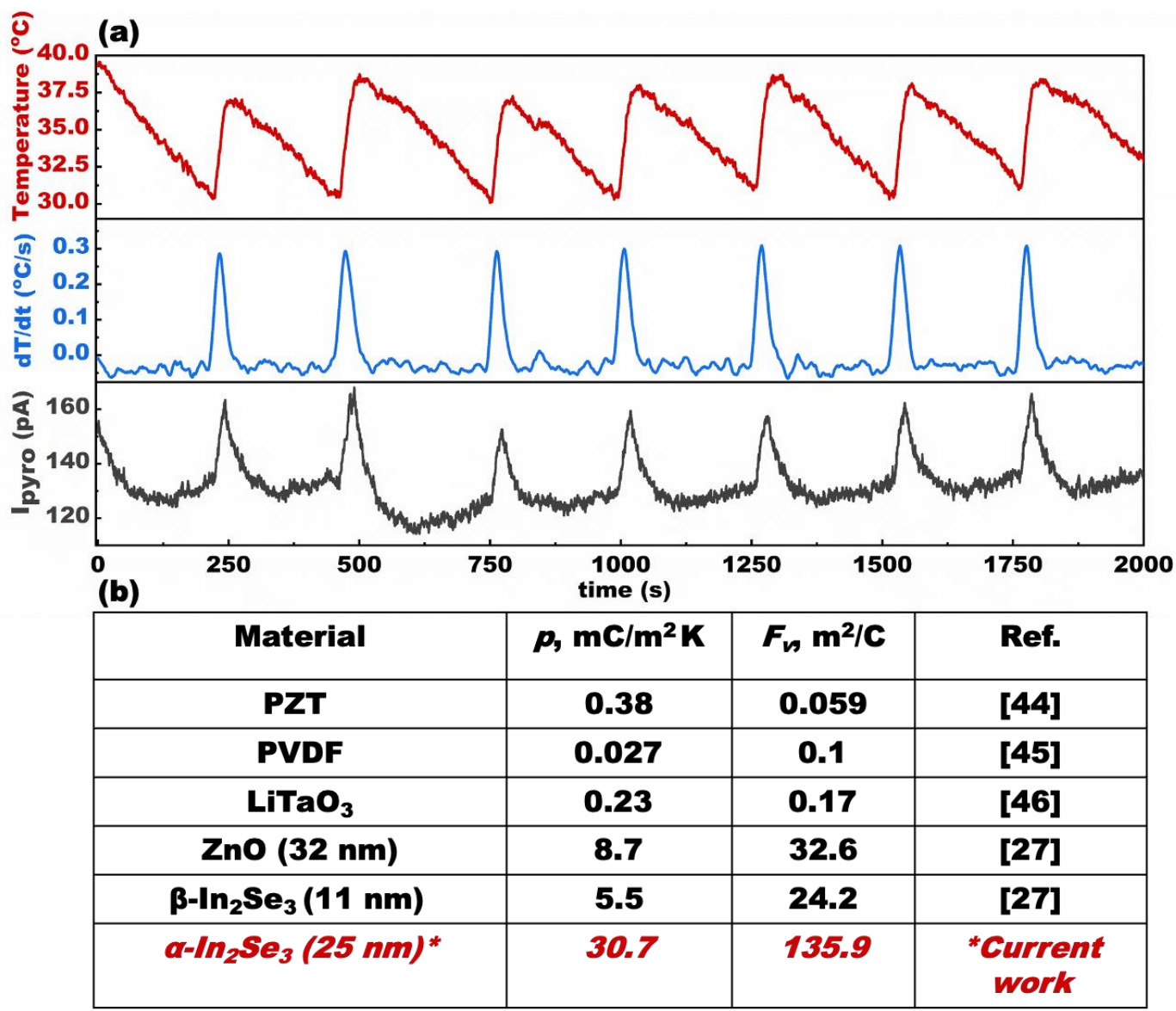

*Figure 2* – **(a)** Direct pyroelectric measurement at ambient conditions (in dark). The data in the stack-plot from top to bottom correspond to temperature, temperature derivative, and the measured pyroelectric current. **(b)** – Comparison table of previously reported pyroelectric coefficients and figure of merits of several conventional ferroelectric materials and for $\alpha$-In$_2$Se$_3$ utilized in the current work (red font).

# Figure 3

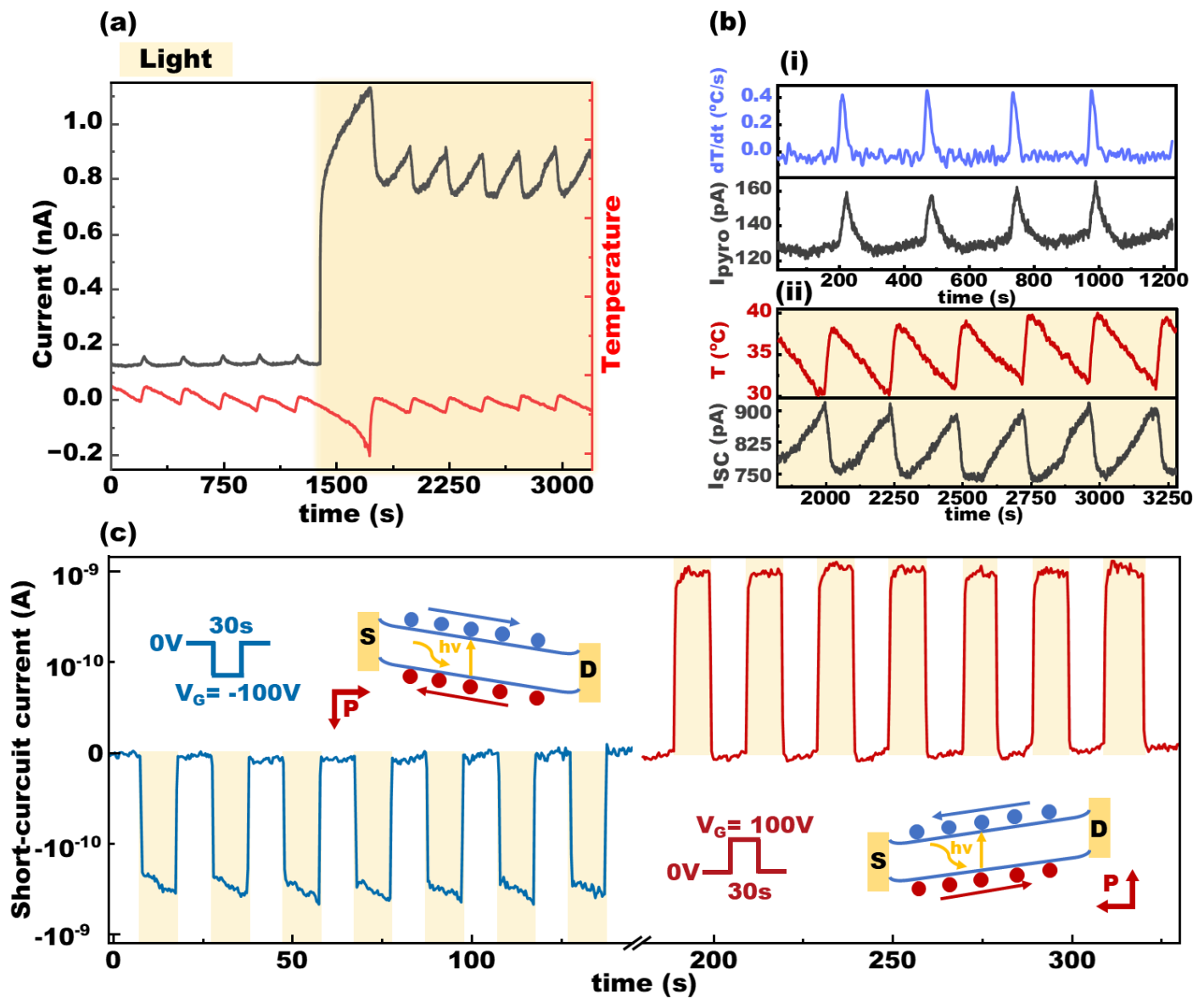

*Figure 3* – **(a)** – Current vs. time measurement under dark and light conditions. The yellow transparent box indicates light illumination. The black and red curves correspond to the measured current and temperature, respectively. **(b)** Zoom in selection from (a) at dark **(i)** and under illumination **(ii)**. **(i)** – corresponds to the pyroelectric current, which follows the temperature-derivative, whereas **(ii)** – corresponds to the short-circuit current under illumination, which monotonously follows the temperature. **(c)** –Short-circuit current after 30-second -100 V (blue) and +100V (red) gate electrode modulation. The insets demonstrate the schematic change of the surface electrostatic potential slope across the device after the gate voltage withdrawal due to the induced IP polarization.

# Figure 4

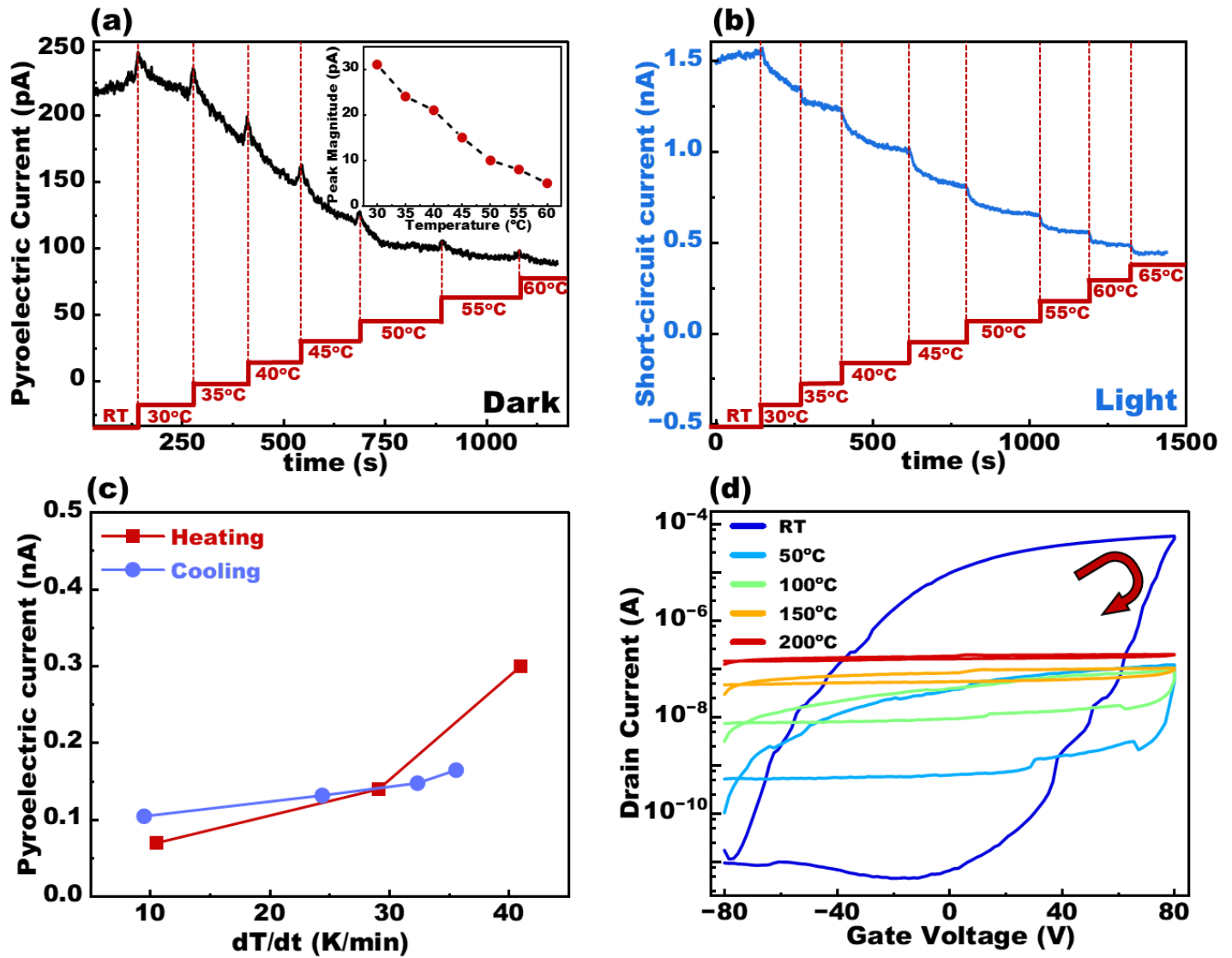

*Figure 4* – Current *vs.* time measurements with gradual temperature growth (red lines) in dark **(a)** and light **(b)** conditions. Under dark conditions, the pyroelectric current decreases with temperature increase. The inset image shows the linear correlation of the pyroelectric current peak with temperature. Under light, the pyroelectric effect is screened by the short-circuit current and reduces with temperature growth. **(c)** – Pyroelectric current *vs.* temperature modulation rate under heating and cooling. **(d)** – Temperature dependence of the transfer characteristics under 1V drain voltage.